# Quantum oscillations of holes in GaN

**Authors**: Chuan F.C. Chang[1]*, Joseph E. Dill[2], Zexuan Zhang[3], Jie-Cheng Chen[4,5], Naomi Pieczulewski[6], Samuel J. Bader[7], Oscar Ayala Valenzuela[8], Scott A. Crooker[8], Fedor F. Balakirev[8], Ross D. McDonald[8], Jimy Encomendero[3], David A. Muller[2,9], Feliciano Giustino[4,5], Debdeep Jena[3,6,9]*, Huili Grace Xing[3,6,9]*

**Affiliations:**
[1]Department of Physics, Cornell University, Ithaca, NY, USA.
[2]School of Applied and Engineering Physics, Cornell University, Ithaca, NY, USA.
[3]School of Electrical and Computer Engineering, Cornell University, Ithaca, NY, USA.
[4]Department of Physics, The University of Texas Austin, Austin, TX, USA.
[5]Oden Institute for Computational Engineering and Sciences, The University of Texas Austin, Austin, TX, USA.
[6]Department of Materials Science and Engineering, Cornell University, Ithaca, NY, USA.
[7]Foundry Technology Research, Intel Corporation, Hillsboro, OR, USA
[8]National High Magnetic Field Laboratory, Los Alamos National Laboratory, Los Alamos, NM, USA.
[9]Kavli Institute at Cornell for Nanoscale Science, Ithaca, NY, USA.
*Corresponding authors: cc2737@cornell.edu, djena@cornell.edu, grace.xing@cornell.edu

**Abstract:** GaN has emerged to be a major semiconductor akin to silicon due to its revolutionary impacts in solid state lighting, critically enabled by p-type doping[1], and high-performance radio-frequency and power electronics[2–4]. Suffering from inefficient hole doping and low hole mobility[5], quantum oscillations in p-type GaN have not been observed, hindering fundamental studies of valence bands and hole transport in GaN. Here, we present the first observation of quantum oscillations of holes in GaN. Shubnikov-de Haas (SdH) oscillations in hole resistivity are observed in a quantum-confined two-dimensional hole gas at a GaN/AlN interface, where polarization-induced doping overcomes thermal freeze-out, and a sharp and clean interface boosts the hole mobility enough to unmask the quantum oscillations. These holes degenerately occupy the light and heavy hole bands of GaN and have record-high mobilities of ~1900 cm$^2$/Vs and ~400 cm$^2$/Vs at 3K, respectively. We use magnetic fields up to 72 T to resolve SdH oscillations of holes from both valence bands to extract their respective sheet densities, quantum scattering times, and the effective masses of light holes (0.5-0.7 m$_0$) and heavy holes (1.9 m$_0$). SdH oscillations of heavy and light holes in GaN constitute a direct metrology of valence bands and open new venues for quantum engineering in this technologically important semiconductor. Like strained silicon transistors, strain-engineering of the valence bands of GaN is predicted[6,7] to dramatically improve hole mobilities by reducing the hole effective mass, a proposal that can now be explored experimentally, particularly in a fully fabricated transistor, using quantum oscillations. Furthermore, the findings of this work suggest a blueprint to create 2D hole gases and observe quantum oscillations of holes in related wide bandgap semiconductors such as SiC and ZnO in which such techniques are not yet possible.

**Main**

The discovery of p-type doping in the wide bandgap semiconductor GaN – to create holes in its valence band as the mobile charge carriers – made blue LEDs and lasers possible[1]. White LEDs have replaced incandescent and fluorescent light bulbs, ushering in an energy-efficient solid-state lighting revolution in photonics. As GaN high-electron mobility transistors (HEMTs) are ushering in a new era of energy-efficient RF and power electronics because of fast switching enabled by high electron mobility[2–4], there is rising interest in p-type GaN transistors for integrated complementary circuits[5,8–11]. Shubnikov-de Haas (SdH) quantum oscillations were observed immediately after the discovery of two-dimensional electron gases (2DEGs) at AlGaN/GaN heterojunctions even before their origin was understood to be due to polarization discontinuity[12]. The dual of the polarization-induced 2DEG, the two-dimensional hole gas (2DHG) in GaN without any impurity doping, was only discovered in 2019[13]. As GaN holes have lower mobility and are harder to access experimentally, fundamental understanding of the 2DHG has lagged that of the 2DEG.

Here, we report successful observation of strong SdH oscillations in such polarization-induced 2D hole gases in GaN. Furthermore, we observe two distinct sets of oscillations due to holes from the light-hole and heavy-hole bands. These observations enable direct metrology of the effective masses of the two bands and offer insights into the valence band structure and transport properties in this semiconductor family.

SdH oscillations – the oscillation in electrical resistance under a magnetic field due to the formation of Landau levels (LLs) – is one of the most powerful and widely used tools to directly probe the band structures and Fermi surfaces of novel materials and novel phases of matter[14–18]. SdH oscillations observed in the technologically relevant semiconductors – Si, Ge, GaAs, and n-type GaN – have helped shape our understanding of their physics by measuring critical material parameters such as the effective mass of charge carriers, Fermi surface geometry, quantum scattering times, gyromagnetic ratio, and spin-orbit coupling[19–22]. The high mobility of charge carriers in these materials, a testament to continual improvements in crystal growth and processing, have enabled the observation of integer and fractional quantum Hall effects when only a few LLs are occupied, revealing a rich palette of geometric and topological effects on the Hall conductance[23–26]. SdH oscillations of holes have been observed in diamond[27] to date, but not in other wide bandgap semiconductors.

Stringent requirements must be met to observe SdH oscillations: mobile carriers and electrical contacts that can survive to cryogenic temperatures and high carrier mobility. First, holes in Mg-doped p-type GaN rapidly freeze out at low temperatures below 130 K[5,28]. This conflicts with the cryogenic nature of SdH experiments due to the requirement that thermal energy must be sufficiently small compared to the energy separation between quantized Landau levels, i.e., $k_B T < \hbar \omega_c$ ($k_B$ is the Boltzmann constant, $T$ is the absolute temperature, $\hbar$ is the reduced Planck's constant, and $\omega_c = eB/m$ is the cyclotron frequency where $e$ is carrier charge, $B$ is the magnetic field, and $m$ is the effective mass). Secondly, the lack of metals with sufficiently large work functions to form ohmic contacts with GaN's deep valence band, lying 7.5 eV below vacuum level, makes electrical contact a challenge even at room temperature[10,29]. Lastly, even if both holes and electrical contacts can survive to cryogenic temperatures, the holes must have sufficiently high mobility to satisfy the condition $\mu_q B > 1$, where $\mu_q$ is the quantum mobility. Under a magnetic field of 25 T, the quantum mobility, which is typically smaller than the Hall or field-effect mobility, must itself be larger than 400 cm$^2$/Vs to observe SdH oscillations.

**High mobility holes persisting to cryogenic temperatures in a GaN/AlN heterostructure**

Figure 1A shows a schematic of the heterostructure we grew by molecular beam epitaxy (MBE) for this study. We measured the magneotransport properties of the metallic 2DHG residing at the interface of wurtzite GaN and AlN[13]. An atomically sharp interface between GaN and AlN is observed in the annular dark field (ADF) scanning transmission electron microscopy (STEM) image of a representative sample in Fig. 1B. Mobile holes are induced electrostatically by a fixed negative sheet charge arising from the polarization discontinuity between GaN and AlN without the need for Mg-doping. The holes thus do not freeze out at cryogenic temperatures. By growing a thin layer of GaN to ensure pseudomorphic compressive strain (by 2.4%) to the single-crystal bulk AlN substrate, a piezoelectric polarization component develops within the GaN layer. The combined discontinuity in spontaneous polarization and piezoelectric polarization determines the 2DHG density. Though Mg-doping is not required to induce the 2DHG[13,30], we employ an epitaxial layer of Mg-doped GaN with annealed Ni/Au contacts for improved conductivity of electrical contacts.

The high quality of the present samples is the result of several key innovations in crystal growth by MBE, including the use of impurity-blocking layers to prevent compensation by n-type impurities[31], the use of bulk AlN single crystals as the starting substrate instead of AlN templates grown on a foreign substrate[32], as well as *in situ* cleaning[33,34] and heterostructure growth techniques[30] specifically developed to take advantage of bulk AlN substrates. Details of crystal growth are included in Methods.

As shown in the energy band diagram (Fig. 1c) of the heterostructure, a triangular quantum well formed by the valence band offset between GaN and AlN confines the polarization-induced holes near the GaN/AlN interface. A self-consistent solution of the 6-band heterostructure k.p equation and the Poisson equation[35] suggests that the Fermi level lies ~ 60 meV below the valence band maximum, creating a degenerate 2DHG in two distinct valence bands – the first heavy hole (HH) subband and the first light hole (LH) subband of the quantum well.

Figure 1d shows that the measured longitudinal resistance $R_{xx}$ of the 2DHG decreases with decreasing temperature at zero magnetic field, confirming that the 2DHG channel is metallic. The positive slope of the Hall resistance $R_{Hall}$ measured up to 9 T confirms that the transport is due to holes (Fig. 1e). At low temperatures, the Hall resistance becomes non-linear with B, which is a consequence of the distinct mobilities of the light and heavy holes[36]. In the presence of such two (light and heavy hole)-band occupation, the density and Hall mobility of carriers in each band must be determined by fitting a two-channel Drude model to the measured nonlinear Hall resistance $R_{Hall}$ and parabolic longitudinal resistance $R_{xx}$. For the present sample, this analysis yields Hall mobilities ~1900 cm$^2$ V$^{-1}$ s$^{-1}$ and ~400 cm$^2$ V$^{-1}$ s$^{-1}$ at 3 K for the light holes and heavy holes, respectively (see Methods). While these mobilities are the highest reported to date to our best knowledge, the condition for the onset of quantum oscillations, $\mu_q B > 1$, involves quantum mobilities, which are typically lower than Hall mobilities. Therefore, a large magnetic field is still required to ensure that quantum oscillations of both holes are clearly resolved.

**Shubnikov-de Haas oscillations of heavy holes and light holes**

Here, using pulsed magnetic fields up to 72 T, we report clear SdH oscillations in $R_{xx}$ with an onset at about 25 T (Fig. 1f). After subtracting a smooth polynomial background, no additional data smoothing is performed (see materials and methods). Characteristic of quantum oscillations, the amplitude grows with magnetic field as the Landau levels become increasingly separated and well-resolved in the vicinity of the Fermi level. Similarly, with decreasing temperature, the sharpening of the Fermi-Dirac distribution relative to the energy gap between neighboring Landau levels (i.e., $k_B T < \hbar\omega_c$) gives rise to an increase in oscillation amplitude. Cooling from 15 K to 1.8 K, we observe another distinct set of oscillations, resolvable only above 50 T. In the following, we show that these small-amplitude, high-frequency oscillations are due to the low-mobility but high-density heavy holes, while the underlying large-amplitude, low-frequency oscillations are due to the higher-mobility, lower-density light holes.

The Onsager relation[14], $A_k = (2\pi e/\hbar)f$, relates each oscillation frequency $f$ to a momentum-space Fermi surface area $A_k$. Figure 2a shows the background-subtracted $R_{xx}$, which is periodic in $1/B$ and remains in phase across temperatures. Their Fast Fourier transform (FFT) spectra in Fig. 2b show a dominant frequency peak at $f_1 = 166 \pm 2$ T observed across temperatures. The oscillatory resistance is fitted to the Lifshitz-Kosevich (LK) formula, $\Delta R_{xx} \propto \exp\left(-\frac{\pi}{\mu_q B}\right)\frac{X}{\sinh X}\cos\left(2\pi f \frac{1}{B}\right)$, where $X = 2\pi^2 \frac{k_B T}{\hbar\omega_c}$. A good fit (dashed lines in Fig. 2a) is obtained for the low-frequency, large-amplitude oscillations. We then subtract out the low-frequency fit to retain only the high-frequency, small-amplitude oscillations, where the characteristic growth in amplitude with increasing field and decreasing temperature is once again observed (Fig. 2c). The FFT spectra of the high-frequency oscillations are displayed in Fig. 2d, revealing a broader peak at $f_2 = 795 \pm 24$ T observed up to 6 K. Relating the momentum-space Fermi surface areas to sheet densities using Luttinger's theorem, $n = 2A_k/(2\pi)^2$, we obtain a sheet density $n_1 = (8.0 \pm 0.1) \times 10^{12}$ cm$^{-2}$ for the low-frequency component, which we attribute to carriers occupying the deeper-lying light hole band with a lower density of states. The high-frequency oscillations, with a corresponding sheet density $n_2 = (3.8 \pm 0.1) \times 10^{13}$ cm$^{-2}$, are due to carriers occupying the heavy hole band that lies on the top of the valence band manifold and has a larger density of states. No temperature dependence of the densities is observed (Fig. 2e), confirming the polarization-induced origin of both light and heavy holes. Though the Hall resistance was not measured beyond 9 T in this study, the relatively low density of the light holes gives it a calculated filling factor $\nu = 4$ at $B = 72$ T (Extended Data Fig. 6), placing it close to the integer quantum Hall regime.

**Direct measurement of heavy and light hole effective masses and quantum scattering times**

The distinct set of oscillations from the light and heavy hole valence bands allows us to extract their individual effective masses. Reports of GaN hole mass in literature scatter from 0.3 to 2.2 $m_0$[37], obtained by optical methods such as excitonic photoluminescence[38–40], optical reflectance[41,42], and spectroscopic ellipsometry[43] in which the light-matter interaction couples the valence to the conduction band. Magnetotransport oscillations measure the valence (or conduction) band effective masses directly without the need to deconvolute the data since there is no interband coupling. However, due to low hole mobility in GaN, direct and unambiguous measurements of hole effective masses by SdH oscillations and cyclotron resonance have not been possible in the past. In addition, SdH oscillations and cyclotron resonance measure effective mass of carriers at the Fermi level, which for degenerate (metallic) carrier concentrations can differ significantly from the zone-center effective mass when the bands are nonparabolic, as is the case for the valence bands of GaN.

Figure 2f shows that the temperature dependence of the light hole and heavy hole FFT amplitudes is described accurately by the thermal damping term in the LK formula. We extract effective mass values of $1.92 \pm 0.16\ m_0$ for the heavy holes and $0.53 \pm 0.01\ m_0$ for the light holes at their respective Fermi wave vectors. The larger mass of the heavy holes gives rise to (i) a larger density of states, consistent with their higher measured density, and (ii) a lower mobility, consistent with the larger magnetic field required to resolve their oscillations.

Knowledge of the effective mass combined with a quantum mobility obtained from the field-dependence of oscillation amplitude gives the quantum scattering time – a measure of disorder broadening due to scattering from interface roughness, charged impurities, and dislocations. We extract a quantum mobility of $368 \pm 14 \text{ cm}^2\text{V}^{-1}\text{s}^{-1}$ for $T = 1.8 \sim 15$ K (Extended Data Fig. 7) for light holes, consistent with the onset of oscillations at about 25 T, and obtain a quantum scattering time of 0.15 ps. For the heavy holes, the relatively small signal-to-noise ratio of the oscillation amplitudes precludes a reliable extraction of quantum mobility using this method; instead, the onset of their oscillations in the approximate range of 50~60 T gives an estimated quantum mobility of $167\sim200 \text{ cm}^2\text{V}^{-1}\text{s}^{-1}$, corresponding to a quantum scattering time of 0.17~0.24 ps. Both quantum mobilities are smaller than the Hall mobilities, with a $\mu_{Hall}/\mu_q$ ratio of ~5 for light holes and 2~2.5 for the heavy holes. As argued for 2DEGs in GaN[44], a $\mu_{Hall}/\mu_q$ ratio in the range of 1.5~9 likely suggests the dominance of a short-range, large-angle scattering mechanism such as interface roughness scattering. If interface roughness scattering is indeed the mobility-limiting mechanism in the present case, there will be room for further improvement in low-temperature mobility with improved roughness interfaces in the future.

**Comparison of measured effective masses with theory reveals a larger-than-predicted light hole mass**

We now compare the hole masses extracted from the experimental quantum oscillations with the those derived from band dispersions obtained using two theoretical methods – the 6-band k.p method and the *ab initio GW* method (see materials and methods). Figure 3a shows the in-plane dispersions of the heavy and light holes. For the k.p method, Burt's exact envelope function theory [45] is used to obtain energy dispersions explicitly taking into account the GaN/AlN heterostructure potential. The energy dispersions shown in Fig. 3a for the k.p method are thus the first heavy and light hole subbands. An anti-crossing of the first light hole subband with the second heavy hole subband (first excited state) in the GaN quantum well produces the kinks at $k \sim \pm 1 \text{ nm}^{-1}$. Separately, we perform state-of-the-art *GW* perturbation theory calculations to obtain *ab initio* band dispersions for a strained GaN bulk crystal (solid lines in Fig. 3a). The kinks in the *ab initio* light hole band results from an anti-crossing with the split-off valence band rather than with higher subbands due to quantization, which are not calculated in the *GW* method.

The effective masses calculated from the dispersions are shown in Fig. 3c along with the experimental values, with $k_F$ calculated from the experimental densities as $k_F = \sqrt{2\pi n}$. Both theoretical methods indicate that significant mixing of the light hole and heavy hole states at the zone center at $k = 0 \text{ nm}^{-1}$ gives them a common effective mass of $0.45 \ m_0$. Away from the zone-center, they recover their individual effective masses of $1.6\sim2.1 \ m_0$ for the heavy hole and $0.29 \ m_0$ for the light hole, until the light hole band anti-crosses with another band and takes on the other band's effective mass. While the measured heavy hole mass of $1.92 \pm 0.16 \ m_0$ is in good agreement with both theoretical methods, the measured light hole mass of $0.53 \pm 0.01 \ m_0$ is substantially larger than the calculated value of $\sim 0.27 \ m_0$ from both methods. In the following, we investigate the magnetic field-dependence of the light hole mass in light of this discrepancy.

**Anomalous field-dependence of the light hole effective mass and extrapolation to $B = 0$ T**

The high magnetic field available in this study and the large signal-to-noise ratio of the light hole oscillations enabled us to perform an LK fit for each light hole oscillation peak (Fig. 4a) to extract the light hole mass as a function of magnetic field. Surprisingly, we observe a linear increase in the light hole mass from 0.48 $m_0$ at 32 T to 0.69 $m_0$ at 72 T (Fig. 4b), with the field-averaged value of 0.53 $m_0$ extracted from FFT amplitudes falling in this range. Extrapolating to $B = 0$ T, we obtain a mass of 0.30 $m_0$, which is close to the calculated value from both the k.p and $GW$ theoretical methods. We currently do not have a clear justification for the linear extrapolation and the origin of the magnetic field-dependence of the light hole mass. A field-dependent effective mass as measured by SdH oscillations has been reported in several materials[46–53]. A linear dependence of the electron mass on magnetic field has been reported in AlGaN/GaN 2DEG heterostructures and attributed to conduction band nonparabolicity[46–48]. We investigated the effects of nonparabolicity by a numerical calculation of the light hole and heavy hole Landau levels from their nonparabolic dispersions (see Methods) and find that for both light and heavy holes, the Landau levels near the Fermi level are evenly spaced and have approximately equal cyclotron effective masses, even out to $B = 32$ T and $B = 72$ T (Extended Data Fig. 6). Consistently, the oscillation frequency is observed to show no dependence on magnetic field (Extended Data Fig. 4). Thus, nonparabolicity alone is unable to explain the field dependence of the light hole mass observed here. A field-dependent hole mass has also been observed in similarly asymmetric GaAs/AlGaAs heterostructures with its origin left unexplained[49,50], as well as in a symmetric double quantum well with its origin attributed to electron-electron interaction[51]. In the present case, further experiments and theory are necessary to corroborate our observed trend and elucidate its origin. We therefore leave it as an open question.

**Conclusion and outlook**

Using polarization-inducing doping to create holes that survive to cryogenic temperatures and a sharp and clean GaN/AlN interface grown by MBE to enable hole mobilities of ~1900 cm$^2$ V$^{-1}$ s$^{-1}$ for light holes and ~400 cm$^2$ V$^{-1}$ s$^{-1}$ for heavy holes at 3 K in two valence bands of GaN, we observe their distinct SdH oscillations, and obtain their respective sheet densities, quantum scattering times, and effective masses. The availability of quantum oscillation as a tool to directly access the Fermi surface and probe the valence bands of GaN opens the door to more fundamental studies of this important semiconductor material family. Of immediate high interest is the response of the valence bands of GaN to applied strain with the important goal to lower the hole effective mass and consequently boost the hole mobility as recently proposed[6,7]. A similar strategy is now successfully employed in commercial p-type silicon transistors. Strain-engineered p-GaN is both promising and necessary to close the performance gap between p- and n-channel GaN transistors, essential for realizing well-balanced GaN CMOS.

On the fundamental side, the light holes in the present study have a low density of $\sim 8 \times 10^{12}$ cm$^{-2}$ and the Fermi level populates only four Landau levels at $B = 72$ T (filling factor $\nu = 4$). Reducing the light hole density by gating or by decreasing the thickness of the GaN layer can push the 2DHG further into the integer quantum Hall regime and may even bring the fractional quantum Hall regime within reach. Such experiments will offer an exciting new platform for studying quantum Hall physics due to the strong coupling of GaN holes to acoustic phonons[6,7], interaction between heavy holes and light holes, and a larger Fermi surface than conventional 2D systems. The strong hole-acoustic phonon coupling and the large density of states at the Fermi level enabled by the large heavy hole mass are conducive to Cooper pairing[54,55] and therefore are of high interest in the future for possible superconductivity in this material platform.

The use of polarization-induced doping to enable quantum oscillation should be applied to other wide bandgap semiconductors in which other methods to create mobile carriers are problematic. For example, in ZnO/MgZnO heterostructures, polarization-doped 2DEGs with $\mu > 10^6$ cm$^2$V$^{-1}$s$^{-1}$ have enabled SdH studies of quantum scattering times[56] and spin-orbit coupling[57] and have entered the integer and fractional Hall regimes[15,58,59]. For SiC, polarization-doped 2DEG in a 3C/4H-SiC heterostructure has enabled an electron mobility of $\mu > 7000$ cm$^2$V$^{-1}$s$^{-1}$ at 32 K[60], and quantum oscillations studies, though not yet reported to our knowledge, are already within reach. For both wide-bandgap ZnO and SiC, the p-type dual – the 2DHG – has not been confirmed to date. Like in GaN, improvements in impurity-control[31] and crystal growth[30] to enable high-mobility uncompensated 2DHG by polarization doping will allow quantum oscillation studies to reveal rich information about these related wide bandgap semiconductor families and harness their full potential.

**Acknowledgments:** We thank Vladimir Protasenko for help with the operation of the molecular beam epitaxy system. We thank Erich Mueller, Neil Harrison, John Singleton, Johanna Palmstrom, John Singleton, Laurel Winter, Changkai Yu, and Reet Chaudhuri for fruitful discussions.

**Funding:** This work was supported in part by SUPREME, one of seven centers in JUMP 2.0, Semiconductor Research Corporation (SRC) program sponsored by DARPA. Characterizations and measurements were performed in part at Cornell NanoScale Facility, a National Nanotechnology Coordinated Infrastructure (NNCI) member supported by NSF Grant No. NNCI-2025233. We also made use of the Cornell Center for Materials Research Shared Facilities (DMR-1719875). The Thermo Fisher Spectra 300 X-CFEG was acquired with support from PARADIM, an NSF MIP (DMR-2039380) and Cornell University. The National High Magnetic Field Laboratory is supported by the National Science Foundation through NSF/DMR-2128556, the State of Florida, and the U.S. Department of Energy.

**Author contributions:** H.G.X., and D.J. supervised the project. C.F.C.C. and Z.Z. grew the GaN/AlN epitaxial structures. J.E.D. and J.E. fabricated the Hall bar devices. C.F.C.C., J.E.D., O.A.V., F.F.B., and S.A.C. performed the 72 T magnetotransport experiments. C.F.C.C. performed the 9 T magnetotransport experiments. N.P. and D.A.M. provided STEM images. C.F.C.C. analyzed the magnetotransport data. J.C. and F.G. performed first-principle GW calculations. C.F.C.C performed other theoretical calculations. C.F.C.C. wrote the manuscript. H.G.X., D.J., S.A.C., and S.J.B. revised the manuscript with inputs from all authors.

**Competing interests:** Authors declare that they have no competing interests.

**Data and materials availability:** Source data supporting the findings of this study will be provided with this paper.

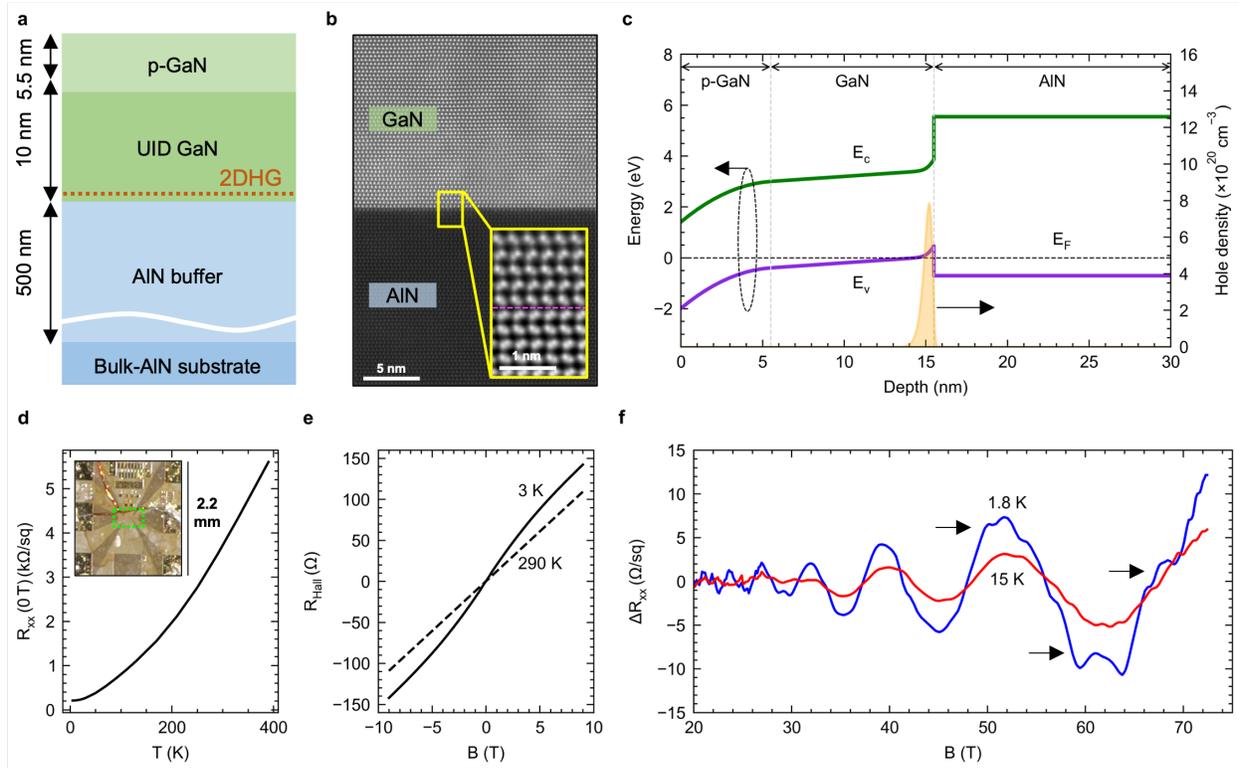

**Fig. 1| Structural and transport properties of a GaN/AlN 2DHG. a**, Schematic of the GaN-on-AlN heterostructure grown by MBE. A 2DHG forms in the unintentionally doped (UID) GaN layer near the GaN/AlN interface. The topmost GaN layer is doped with Mg to reduce contact resistance to the underlying 2DHG. **b**, ADF-STEM image of a representative sample showing an atomically sharp GaN/AlN interface. (Inset) Integrated Differential Phase Contrast (iDPC) image showing the metal-polarity of the heterostructure. **c**, Energy band diagram of the heterostructure showing the conduction band minimum ($E_c$) and the valence band maximum ($E_v$) as a function of depth from the sample surface. A triangular quantum well is formed at the GaN/AlN interface, where the Fermi level ($E_F$) lies below the valence band maximum, creating a degenerate 2DHG. **d**, Longitudinal or sheet ($R_{xx}$) resistance of the 2DHG as a function of temperature, showing clear metallic behavior. (Inset) Photograph of a fabricated Hall bar used for high-field magnetoresistance measurements. The channel region of the Hall bar is indicated in the green box. **e**, Hall resistance $R_{Hall}$ (i.e. $R_{xy}$) measured up to 9 T at 3 K and 290 K indicate p-type charge carriers. Nonlinearity in $R_{Hall}$ develops at low temperatures, suggesting the co-presence of carriers with distinct mobilities. **f**, Longitudinal resistance as a function of magnetic field, measured up to 72 T. A smooth polynomial background is subtracted. Clear quantum oscillations are observed starting around $B = 25$ T. The arrows indicate signatures of an additional oscillation component that emerges at the lowest temperatures and highest fields.

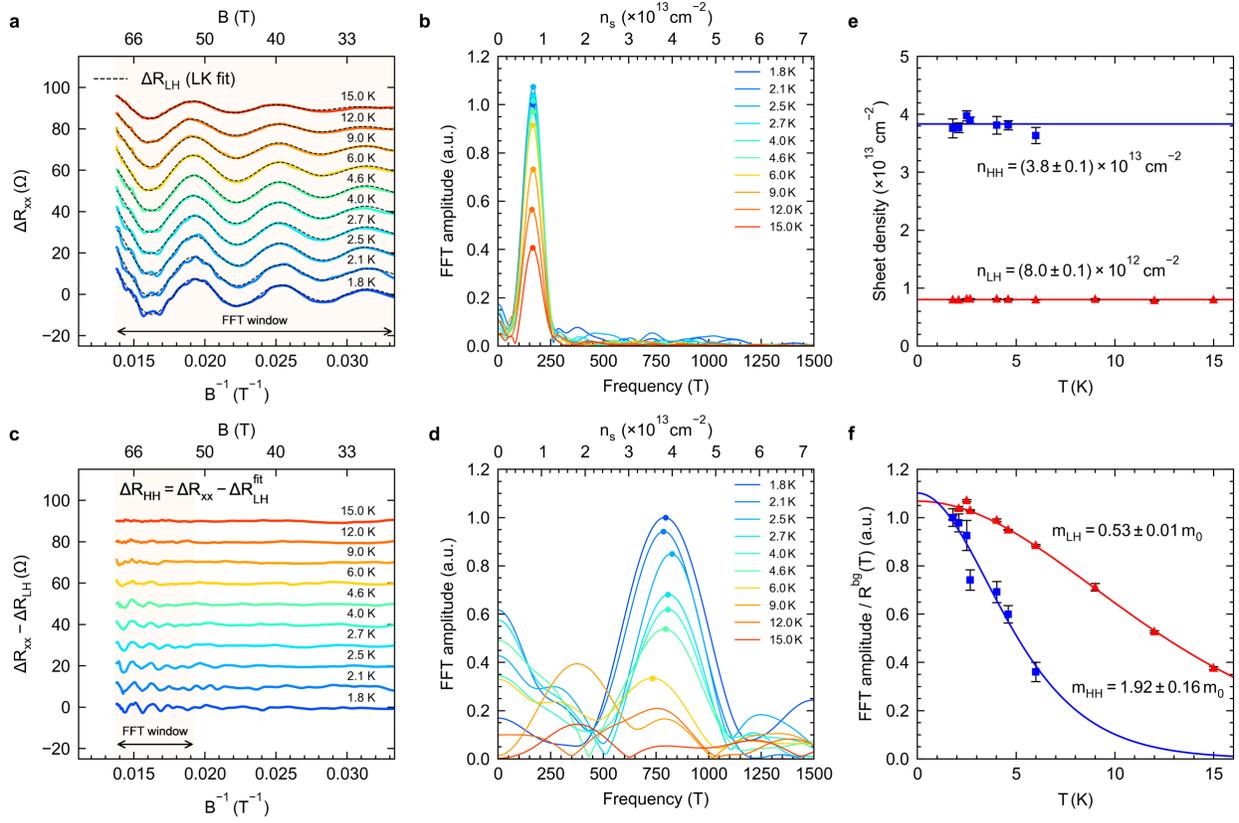

**Fig. 2| FFT analysis of SdH oscillations of 2DHG. a**, The background-subtracted longitudinal resistance $\Delta R_{xx}$ plotted against 1/B, for temperatures from 1.8 K to 15 K. The dashed lines show least-square fits to the single-channel LK formula, performed separately for each temperature. **b**, FFT spectra taken in the magnetic field window indicated in (**a**), normalized to the peak amplitude at 1.8 K. The peaks are indicated as solid circles. A dominant frequency peak at $f_1 = 166 \pm 2$ T is observed across all temperatures. **c,** Small-amplitude, high-frequency oscillations remaining after a subtraction of each curve in (**a**) by its LK fit. **d**, FFT of the small-amplitude, high-frequency oscillations, revealing a broader peak at $f_2 = 795 \pm 24$ T well-resolved up to 6 K. A Hanning window is applied for the FFT spectra presented here to suppress spectral leakage. Extended Data Fig. 3 shows the FFT spectra obtained using different windowing functions. **e**, Sheet density of light (triangles) and heavy holes (squares) obtained from the oscillation frequencies extracted by FFT. The solid lines show the density values averaged across the measured temperatures. **f**, Temperature dependence of the FFT amplitude for both the light hole and heavy hole peaks, first divided by the field-averaged background resistance, $R^{bg}$, at each temperature and then normalized to the value at the base temperature $T_0 = 1.8$ K. Solid lines are fits to the temperature-dependent factor $\alpha\chi/\sinh(\chi)$ of the LK formula where $\alpha$ is a constant pre-factor. In (**e**) and (**f**), the plotted values and error bars indicate the mean and standard deviation across different FFT window functions used.

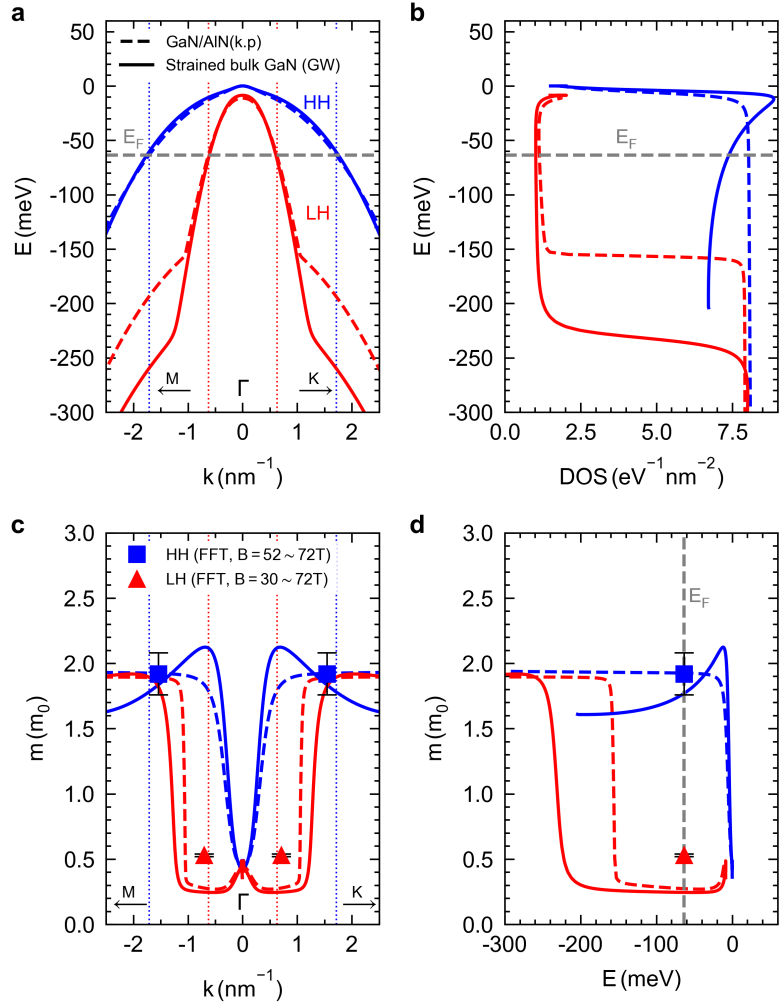

**Fig. 3| Calculated band dispersions and effective masses. a**, Heavy hole and light hole dispersions in the plane of the 2DHG (perpendicular to the c-axis of the wurtzite crystal) calculated using the k.p method for a GaN/AlN quantum well (dashed curves) and using first-principle *GW* calculations for a GaN bulk crystal fully strained to AlN (solid curves). $k = 0$ is Γ-point (zone center); $k < 0$ is the Γ → M direction, and $k > 0$ is the Γ → K direction. The horizontal dashed line is the Fermi level determined from a self-consistent k.p-Poisson calculation. The vertical dashed lines indicate the Fermi wave vector of the light and heavy holes. The kinks at $k \sim \pm 1$ nm$^{-1}$ in the light hole dispersion calculated by the k.p method indicates an anti-crossing with the second heavy hole subband of the quantum well (not shown). For the first-principle light hole dispersion, the kinks indicate an anti-crossing with the split-off valence band (not shown). **b**, Density of states (DOS) per unit area of the light and heavy hole bands as a function of energy. **c-d**, Cyclotron effective masses of the light and heavy holes, calculated from the dispersions in (**a**), as a function of in-plane wave vector (**c**) and energy (**d**). The cyclotron effective mass is directly related to the density of states per unit area via DOS = $m/(\pi\hbar^2)$. The heavy hole and light hole mass determined from FFT amplitudes are shown as squares and triangles, respectively. Error bars indicate one standard error. In (**c**), the Fermi wave vectors of the experimental data points are calculated from the measured densities as $k_F = \sqrt{2\pi n}$.

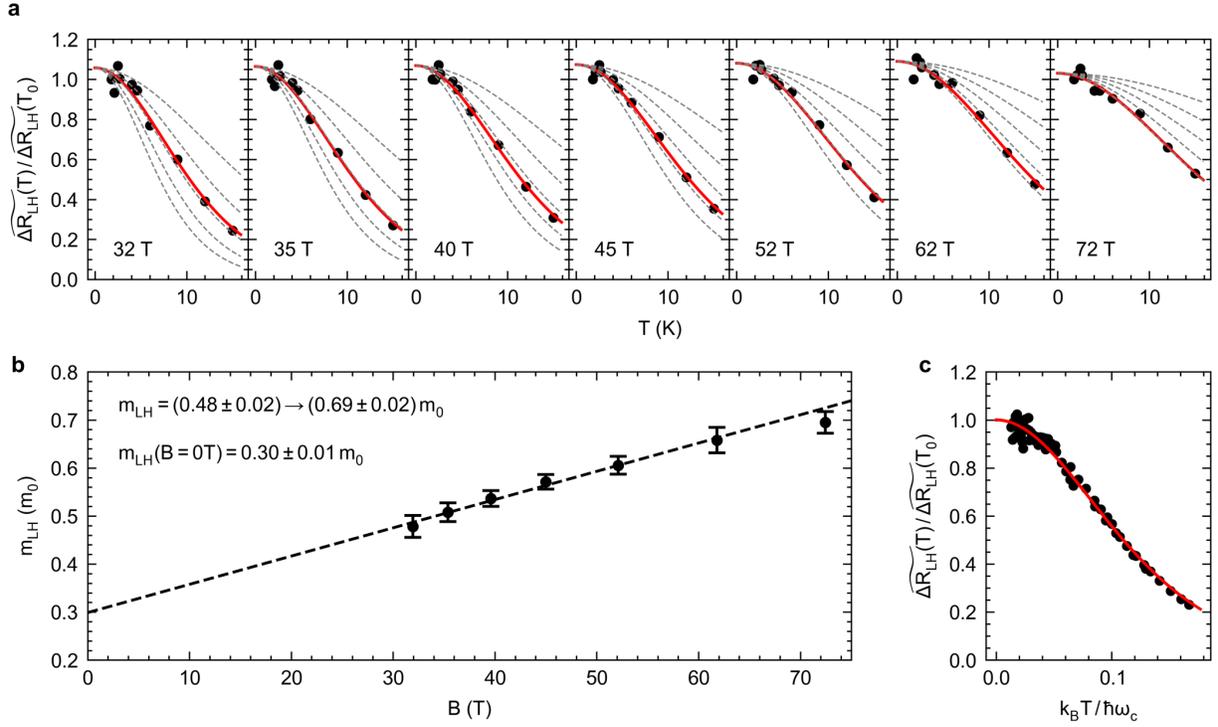

**Fig. 4| Field dependence of light hole mass. a**, Normalized oscillation amplitude taken from the fitted dashed lines in Fig. 2a at various magnetic fields. $\widetilde{\Delta R_{LH}} = \frac{\Delta R_{LH}}{R^{bg}(T)}$ is the background-normalized light hole oscillation amplitude, and $T_0 = 1.8$ K is the base temperature. At each field, the effective mass is independently obtained by fitting to the $\alpha\chi/\sinh(\chi)$ term of the LK formula where $\alpha$ is a constant pre-factor. The grey dashed lines show the LK curves for $0.3\ m_0 \to 0.7\ m_0$ from top to bottom. **b**, Light hole effective mass as a function of magnetic field, showing a linear increase from $0.48 \pm 0.02\ m_0$ at $B = 32$ T to $0.69 \pm 0.02\ m_0$ at $B = 72$ T. The dashed line is a linear fit with a slope of $0.006\ m_0$/T and extrapolates to $0.30\ m_0$ at $B = 0$ T. **c,** The same data in (**a**), scaled by the fitted $\alpha$ and plotted against $k_B T/\hbar\omega_c$, where the fitted field-dependent $m_{LH}$ is used to compute $\omega_c = eB/m_{LH}$. In this scaling, the use of a field-dependent $m_{LH}$ instead of a one field-independent value collapses the data points in (**a**) into a single curve well-described by the LK formula.

# Methods

**Sample Preparation**

GaN/AlN heterostructures were epitaxially grown in a Veeco Gen10 molecular beam epitaxy (MBE) system. Metal (Al, Ga, Mg) fluxes were provided by effusion cells containing high-purity elemental sources. Active nitrogen flux was provided by high-purity $N_2$ gas flowing at a rate of 0.3 sccm through an RF plasma source operating at a power of 400 W. The resulting growth rate was 0.21 µm/hr for all layers in the heterostructure. In situ monitoring of film growth was performed using reflection high-energy electron diffraction (RHEED) apparatus by KSA Instruments with a Staib electron gun operating at 14.5 kV and 1.45 A. The single-crystal, Al-polar AlN substrates used in this study were from Crystal IS and have dislocation density $< 10^4/cm^2$ [32]. The substrates were initially cleaned *ex situ* by solvents (acetone, isopropanol), acids (phosphoric acid, sulfuric acid, and hydrofluoric acid) and de-ionized water, as outlined in our previous works[33,34]. They were then mounted on faceplates and loaded into the load-lock chamber of the MBE system where they were outgassed at 200°C for 7 hours. Prior to the epitaxial growth, the substrates were cleaned *in situ* by repeated cycles of Al adsorption and desorption to remove native surface oxide[33,34]. A 500 nm AlN buffer layer was grown under Al-rich conditions at a thermocouple temperature of $T_{TC} = 1060°C$, after which excess Al was thermally desorbed at $T_{TC} = 1100°C$ for 45 minutes. Then, ten periods of thin $Al_{0.95}Ga_{0.5}N$ layers (2-3 monolayers in each), separated by 25 nm AlN spacers, were grown at $T_{TC} = 875°C$. These layers serve to prevent n-type silicon impurities floating on the growth front during AlN buffer growth from reaching the GaN/AlN interface and compensating the 2DHG[31]. The substrate was cooled down to $T_{TC} = 835°C$ before starting GaN growth. Then, the 15 nm GaN layer was grown without interruption at the GaN/AlN interface. The last 5 nm of the GaN layer is heavily doped with Mg to enable low-resistance metal contacts with the 2DHG. Prior to this step, the Mg source shutter was kept closed. After epitaxial growth, the substrates were immediately cooled down to room temperature, and excess Ga droplets remaining on the surface were removed by HCl. The samples were fabricated into Hall bars using standard photolithography and annealed Ni/Au contacts. A photograph of a fabricated Hall bar is displayed in the inset of Fig. 1d of the main text.

**Magnetotransport measurement**

Low-field ($\leq 9$ T) magnetotransport measurements were performed on the fabricated Hall bars in a Dynacool Physical Property Measurement System (PPMS) system from Quantum Design using the standard six-probe technique with a DC excitation current of 50 µA. Longitudinal ($R_{xx}$) and Hall ($R_{xy}$ or $R_{Hall}$) resistances are symmetrized and anti-symmetrized, respectively, in magnetic field. Pulsed-field measurements were performed in the 75 T Duplex magnet at the National High Magnetic Field Laboratory Pulsed Field Facility at Los Alamos National Laboratory. Temperatures down to 1.8 K were obtained using a $^4$He cryostat with samples immersed in superfluid $^4$He for optimal heat-sinking. Copper wires were attached to the Ni/Au contact pads of the Hall bar using silver epoxy and secured using GE varnish to minimize mechanical vibrations during pulsed field measurements. To minimize induced emf produced by the pulsed field, care was taken to minimize any open-loop area formed by the copper wires. To obtain the cleanest SdH signals possible, Hall resistance was not measured in the 72 T pulsed measurements to eliminate noise contributed by additional wiring. Four-probe measurements of longitudinal resistance of the Hall bar $R_{xx}$ were performed using an AC modulation method, with phase-sensitive detection of the signal performed off-line using a software implementation of a lock-in amplifier. An AC modulation frequency of 40 kHz and an RMS excitation current of 50 µA were used. The phase factor used for lock-in processing is chosen to give maximum oscillation amplitudes in the in-phase component of the measured voltage, which is then used to obtain resistance. We find that the oscillation amplitudes are nearly independent of the phase factor in a $\pm 10°$ window. The phase factor can be adjusted in this narrow window to obtain approximately equal resistances at $B = 0$ T for all temperatures since DC measurements at $B = 0$ T suggest that the sheet resistance flattens below ~20 K. Even though the out-of-phase component cannot be completely removed at all fields due to the high modulation frequency and the finite RC time constant across the sample, such a choice of phase factor always makes the out-of-phase component featureless, i.e., containing no oscillatory signals, and flat with magnetic field. The time constant used for digital lock-in processing is fixed at 250 µs, which is ten times the AC modulation period. All data presented in this work did not undergo additional smoothing (e.g., low-pass or Savitzky-Golay filtering) which may artificially dampen oscillations.

Though commonly neglected, strictly speaking, it is not $\Delta R_{xx}$ vs. T that should be fitted to Lifshitz-Kosevich (LK) formula to obtain effective mass, but rather $\Delta R_{xx}/R^{bg}$ vs. T, where $R^{bg}$ is the background, non-oscillatory longitudinal magnetoresistance, which is distorted in AC measurements. Normalizing by $R^{bg}$ would affect the value of effective mass only if $R^{bg}$ is temperature-dependent. Adhering to this procedure to make sure that $R^{bg}$ up to 72 T is not strongly temperature-dependent, we approximated $R^{bg}$ from DC-current measurements. DC measurements require two separate measurements at each temperature – with magnetic field pointing upward and downward – to average out the distortion by induced emf due to the large $dB/dt$ in pulsed field shots. Additionally, DC measurements in a pulsed-field environment are inherently noisier, particularly when two datasets (upward field and downward field) are averaged, leading to washed-out, unresolved SdH oscillations. Nevertheless, these measurements were performed only to approximate $R^{bg}$ and check its temperature dependence. We perform two sets of DC measurements at 1.4 K and 17 K and interpolate between them. Between 32 T and 72 T, $R^{bg}$ at 17 K is about 10% larger than at 1.4 K.

**Simulation**

Energy band diagrams and band dispersions of the quantum well heterostructure were simulated by self-consistently solving the 6-band k.p equation with the Poisson equation using the commercial software `nextnano`[35]. The Rashba-Sheka-Pikus parameters ($A_{1\sim6}$) and the deformation potentials ($a_{cz} - D_1, a_{ct} - D_2, D_{3\sim6}$) are taken from ref. 61 and ref. 62, respectively. In addition, Burt's exact envelope function theory[45] is used to generalize the k.p method to heterostructures and obtain the band structures for the quantum-confined 2DHG in a strained GaN/AlN quantum well. The spontaneous polarization constants $P_{sp}$ and piezoelectric polarization constants $e_{33}, e_{31}, e_{51}$ are taken from ref. 63. Other material parameters are taken from ref. 64. All relevant input parameters used in k.p-Poisson calculations are listed in Table S1.

*Ab initio* band structure calculations are performed within the framework of density functional theory (DFT) and GW perturbation theory using the Quantum ESPRESSO package[65] and the BerkeleyGW code[66]. The Wannier90 code[67] is utilized to interpolate GW quasiparticle energy. The local density approximation (LDA) to the exchange and correlation functional is used[68], and the spin-orbit coupling is included through fully-relativistic optimized norm-conserving Vanderbilt pseudopotentials[69]. To increase accuracy, we include Ga semicore states in the pseudopotential. A plane-wave kinetic energy cutoff of 180 Ry is used for Kohn-Sham wavefunctions. The Brillouin zone is sampled on a 10×10×10 Γ-centered k-mesh. The effect of strain due to growth on the substrate AlN is considered by epitaxially straining bulk GaN to the lattice of AlN in the basal plane. We use the experimental lattice constants of GaN and AlN instead of *ab initio* values[61]. We perform one-shot GW calculations to obtain the quasiparticle renormalization of band energies[70]. We use 800 spinor states to converge quasiparticle eigenvalue corrections, and a kinetic energy cutoff of 15 Ry for the dielectric matrix. Both DFT and GW eigenvalues are averaged over spin doublets and axial symmetry around z is imposed to obtain isotropic dispersions in the plane. For the Wannier interpolations of GW eigenvalues, maximally localized Wannier functions are obtained from N-2p initial projections for the 12 spinor bands in the highest disentangled valence band manifold, and the coarse grid is also a 10×10×10 Γ-centered k-mesh.

**Transmission Electron Microscopy**

Cross-section lamella of a representative GaN/AlN heterostructure sample was prepared using the Thermo Fisher Helios G4 UX Focused Ion Beam. The lamella was prepared with a final milling step of 5 keV to reduce damage. Scanning transmission electron microscopy (STEM) measurements were taken with an aberration-corrected Thermo Fisher Spectra 300 CFEG operated at 300 keV.

**Background magnetoresistance**

The combined light hole and heavy hole oscillations are obtained by subtracting out a smooth polynomial background from the measured longitudinal resistance (Extended Data Fig. 1a). The amplitudes of light hole oscillations are determined by fitting the background-subtracted longitudinal resistance to the LK equation for a single conducting channel (Extended Data Fig. 1b and Fig. 2a). Heavy hole oscillations are then obtained by subtracting out the fitted light hole oscillations (Fig. 2c). The negative magnetoresistance of the background is an artifact of the RC time constant of the circuit and the high AC modulation frequency required for pulsed-field measurements. It is found to be dependent on the pulsed magnet used and the electrical leads attached to the sample. As detailed in the Materials and Methods section above, the background magnetoresistance $R^{bg}$, which is required to normalize $\Delta R_{xx}$ before extracting effective mass, is approximated by DC-current measurements.

On the other hand, extracting band-resolved mobilities requires accurate measurements of the absolute longitudinal and Hall resistances, which are fitted to a classical two-channel Drude model. However, this method does not require the large pulsed magnetic field. Extended Data Fig. 2 shows low-field ($\leq 9$ T) measurements using DC current and static magnetic field. Longitudinal and Hall resistances have been symmetrized and anti-symmetrized, respectively, as a function of magnetic field. The characteristic parabolic longitudinal resistance and nonlinear Hall resistance are indeed observed. Using the fitting algorithm described in our recent work[36], we obtain transport mobilities $\mu_{LH} = 1858{\sim}1986 \text{ cm}^2/\text{Vs}$ and $\mu_{HH} = 381{\sim}464 \text{ cm}^2/\text{Vs}$ at $T = 3$ K, and $\mu_{LH} = 1537{\sim}1657 \text{ cm}^2/\text{Vs}$ and $\mu_{HH} = 376{\sim}438 \text{ cm}^2/\text{Vs}$ at $T = 20$ K.

Fast Fourier transform (FFT) is used to determine the oscillation frequencies of the light and heavy holes and their field-averaged amplitudes. The finite magnetic field window in which FFT is performed leads to spectral leakage in the resulting spectrum, which is mitigated by multiplying the data with a window function prior to FFT. The resulting spectrum may depend on the specific window chosen. Extended Data Fig. 3 shows FFT spectra obtained when using different window functions. The error bars in Figs. 3A and 3B reflect the spread (one standard deviation) in FFT peak frequencies and amplitudes when different window functions are applied.

FFT analysis is a standard approach for analyzing quantum oscillations. However, the resulting peak frequency and amplitude reflect an average over the magnetic field range in which FFT is performed. This masks potential field dependence in the carrier density and mass. As described in the main text, a field-dependent light hole mass is revealed when tracking the temperature dependence of each oscillation extremum (Fig. 4). Similarly, we can obtain the oscillation frequency near each field from spacing of the oscillation extrema. When plotted in $1/B$, the spacing between one extremum to the next gives the half oscillation period, $\frac{1}{2}T_{B^{-1}}$, which can then be converted into sheet density via $n = \frac{2e}{h}f = \frac{2e}{h}\frac{1}{T_{B^{-1}}}$. As shown in Extended Data Fig. 4, no field dependence of the light hole density is found. This is also consistent with the narrow frequency peaks for the light holes in their FFT spectra (Fig. 2 and Extended Data Fig. 3).

**Definition of cyclotron mass**

In the semiclassical description, an external magnetic field leads to orbital motion of electrons in both real and momentum space in the plane perpendicular to the external field. For electrons occupying different available energy levels, the field causes electrons to orbit on different loci of constant energy in momentum space. The enclosed momentum-space area is a function of energy, which we denote as $A(E)$. By the dispersion relation $E(k)$, the momentum-space area is also a function of the wave vector $k$, so it can also be written as $A(k)$. For each orbit, the angular frequency is $\omega_c = \frac{eB}{m_{CR}}$, where the cyclotron effective mass $m_{CR}$ is defined as

$$m_{CR} \equiv \frac{\hbar^2}{2\pi}\frac{\partial A(E)}{\partial E} = \frac{\hbar^2}{2\pi}\frac{\partial A}{\partial k}\left(\frac{\partial E}{\partial k}\right)^{-1}.$$

For a 2D system that is approximately isotropic in the plane perpendicular to the field, we have a circular Fermi surface in momentum space where $A(k) = \pi k^2$, and the cyclotron mass becomes

$$m_{CR}(k) = \hbar^2 k \left(\frac{\partial E}{\partial k}\right)^{-1}$$

The cyclotron mass is related to the first derivative of $E(k)$ and is directly related to the density of states per unit area $\mathrm{DOS}(E)$ via

$$\mathrm{DOS}(E) = \frac{m_{CR}(E)}{\pi\hbar^2}$$

$m_{CR}$ is in general distinct from the other commonly encountered "curvature" effective mass $m_{curv} \equiv \hbar^2 \left(\frac{\partial^2 E}{\partial k^2}\right)^{-1}$, which is related to the second derivative of $E(k)$. For a parabolic band described by $E(k) = \hbar^2 k^2/2\widetilde{m}$, $m_{CR} = m_{curv} = \widetilde{m}$. For nonparabolic bands, $m_{CR}(k)$ and $m_{curv}(k)$ are approximately equal at those $k$ points where the dispersion is locally well-described by a parabola. The difference between $m_{CR}$ and $m_{curv}$ is illustrated in Extended Data Fig. 4 for the light hole and heavy hole dispersions of the GaN/AlN quantum well. This distinction is necessary for a fair comparison of the measured masses with those derived from theoretical band dispersions.

**Landau levels for nonparabolic dispersions**

In the quantum mechanical description of cyclotron motion[14], the Bohr-Sommerfeld rule leads to a quantization of the real-space area of cyclotron orbits such that they contain integer multiples of the flux quantum, $\phi_0 = \hbar/e$. The momentum-space area is also quantized as

$$A_n = \left(n + \frac{1}{2}\right)\frac{2\pi eB}{\hbar} = A_n(B), \quad n = 0, 1, 2, \ldots$$

For a circular Fermi surface where $A = \pi k^2$, a quantization of the momentum-space area implies a quantization of allowed wave vectors

$$k_n(B) = \sqrt{\frac{A_n(B)}{\pi}} = \sqrt{\left(n+\frac{1}{2}\right)\frac{2eB}{\hbar}}, \quad n = 0, 1, 2, ...$$

which in turn leads to quantized energy levels. Lifting of spin-degeneracy by Zeeman effect would split each $n$ into two levels, $n\uparrow$ and $n\downarrow$. Each $n\uparrow\downarrow$ is a Landau level (LL), labeled by $N$. Note that contrary to the common derivation of LLs by the replacement $\vec{p} \to \vec{p} + e\vec{A}$ in an effective-mass Hamiltonian, this derivation does not assume a parabolic dispersion. Consequently, the LLs may not be equally spaced in the case of nonparabolic dispersions. The LLs are obtained by first calculating the quantized wave vectors and then calculating their corresponding energies from the dispersion. Furthermore, each LL may have a different cyclotron frequency and cyclotron mass. This calculation is illustrated in Extended Data Fig. 6 for $B = 32$ T and $B = 72$ T, neglecting Zeeman-splitting so that each of the plotted energy levels (horizontal lines) contains two spin-degenerate LLs. At $B = 32$ T, eight light hole LLs are occupied; at $B = 72$ T, four light hole LLs are occupied. Nonparabolicity of the light and heavy hole bands is captured by the non-linearity of the Landau levels as a function of magnetic field, most evident in low fields (Extended Data Fig. 6e).

For electrical transport, only charge carriers near $E_F$ are relevant. Despite the nonparabolicity of both bands, the cyclotron masses of the LLs in the vicinity of $E_F$ are approximately constant (0.27 ~ 0.28 $m_0$ for the light holes and 1.93 $m_0$ for heavy holes), at both $B = 32$ T and $B = 72$ T. Consistently, the LLs for both the light and heavy holes are approximately equally spaced near $E_F$ and are linear in magnetic field, i.e.,

$$E_n(B) \approx \hbar \frac{eB}{m_{CR}^{E_F}}\left(n+\frac{1}{2}\right)$$

where $m_{CR}^{E_F}$ is the cyclotron mass of the LLs near $E_F$. This is confirmed by the field independence of the light hole density shown in Extended Data Fig. 4.

**The Lifshitz-Kosevich formalism for quantum oscillations**

When the Fermi level (defined here to be synonymous with the chemical potential), is approximately fixed in an experiment, the density of states (DOS) at $E_F$ oscillates with increasing magnetic field as the LLs separate from each other, crossing $E_F$ one-by-one. Quantities that depend on the DOS at $E_F$, such as electrical resistance, will exhibit oscillations accordingly. The oscillation amplitude depends on the magnitude of change of the DOS at $E_F$ as a function of magnetic field, dampened by the finite width of the LLs and the finite sharpness of the Fermi-Dirac distribution around $E_F$. In the Lifshitz-Kosevich formalism, the oscillatory component of longitudinal resistance ($R_{xx}$) has the form

$$\Delta R_{xx} = C R_D R_T \cos\left(2\pi f \frac{1}{B}\right)$$

$$R_D = \exp\left(-\frac{\pi}{\omega_c^{E_F} \tau_q}\right) = \exp\left(-\frac{\pi}{\mu_q B}\right)$$

$$R_T \equiv \frac{\chi}{\sinh \chi}$$

where $C$ is an amplitude pre-factor, $f$ is the oscillation frequency in $B^{-1}$, $\omega_c^{E_F} = eB/m_{CR}^{E_F}$ is the cyclotron frequency of the LLs near $E_F$, $\tau_q$ is the quantum scattering time, $\mu_q \equiv e\tau_q/m_{CR}^{E_F}$ is the quantum mobility, and $\chi \equiv 2\pi^2 \frac{k_B T}{\hbar \omega_c}$.

The Onsager relation[14] relates $f$ to the momentum-space area via $A = \frac{2\pi e}{\hbar} f$. For an isotropic dispersion with a circular Fermi surface, $A$ is related to the sheet density via Luttinger's theorem, $n = A_k/(2\pi^2)$, and thus $f$ gives a direct measurement of the sheet density via $n = \frac{2e}{h} f$. The Dingle factor $R_D$ describes the reduction of oscillation amplitude by disorder, which for a scattering time of $\tau_q$, leads to a broadening of the LLs by $\hbar/\tau_q$ via the uncertainty principle. This disorder-damping of the oscillation amplitude is especially severe if the LLs are closely spaced near $E_F$ – when $\hbar\omega_c^{E_F}$ is small due to a large $m_{CR}^{E_F}$ or small $B$ – thus the dependence on $\omega_c^{E_F}$ in the Dingle factor. Similarly, the thermal damping factor $R_T$ describes the reduction of oscillation amplitude due to the broadening of the Fermi-Dirac distribution around $E_F$ at finite temperatures. This reduction factor is a function of the ratio of thermal energy $k_B T$ to the near-$E_F$ spacing between LLs, $\hbar\omega_c^{E_F}$.

As shown in the previous section despite the nonparabolicity of the bands, the cyclotron masses of the LLs nearest to $E_F$ are approximately constant (equal to a single value of $m_{CR}^{E_F}$), and the LLs are approximately equally spaced by $\hbar\omega_c^{E_F}$. To determine $m_{CR}^{E_F}$, the standard procedure is to analyze the temperature dependence of the oscillation amplitude at a particular field. The standard approach is to first normalize the amplitudes by the amplitude at the lowest measured temperature and then fit the temperature dependence to $\chi/\sinh\chi$ with $m_{CR}^{E_F}$ as the only fitting parameter. Since $\chi/\sinh\chi$ approaches unity at $T = 0$ K, this approach makes the explicit assumption that the lowest temperature is sufficiently low that the measured amplitude approaches the $T = 0$ K limit. This assumption can be removed by fitting instead to $\alpha\chi/\sinh\chi$ where $\alpha$ is a constant fit parameter greater than unity. When fitting FFT amplitude, the value of $B$ used in the expressions of $\omega_c$ and $\chi$ is based on the harmonic mean $B^*$ of the FFT window: $\frac{1}{B^*} = \frac{1}{2}\left(\frac{1}{B_{\text{low}}} + \frac{1}{B_{\text{max}}}\right)$.

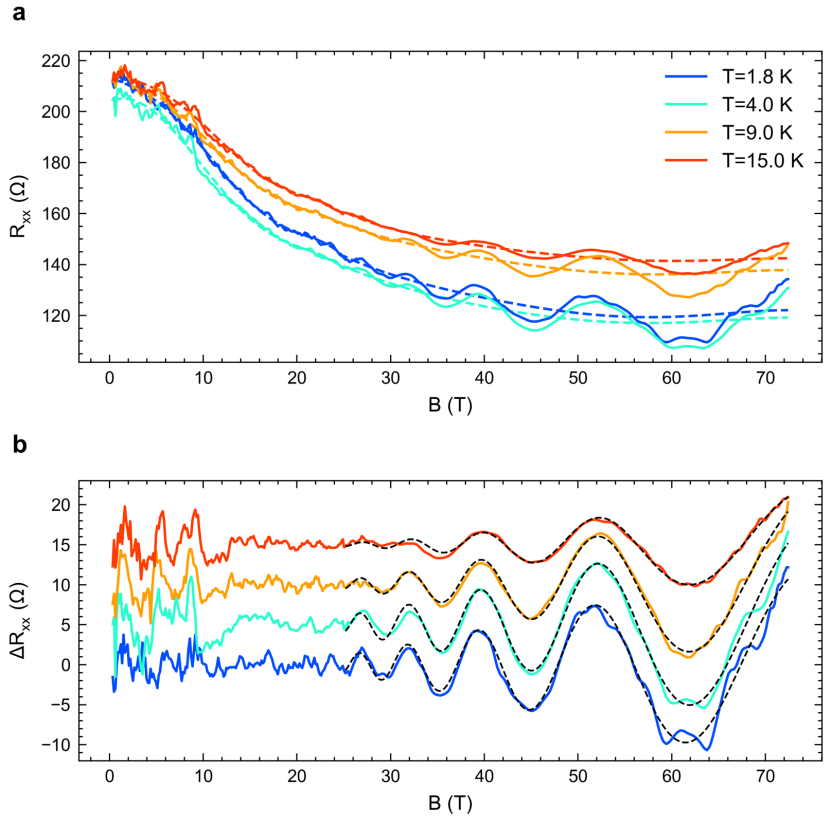

**Extended Data Fig. 1| Background subtraction of longitudinal resistance $R_{xx}$. a**, The solid lines show the measured longitudinal resistance, from which a fifth-order polynomial background (dashed lines) is fitted. **b**, Subtracting out the polynomial background yields the combined light hole and heavy hole oscillations. The combined oscillations are then fitted to the single-channel LK equation (black dashed lines) to obtain the light hole oscillations which, when subtracted, reveals the heavy hole oscillations shown in Fig. 2c.

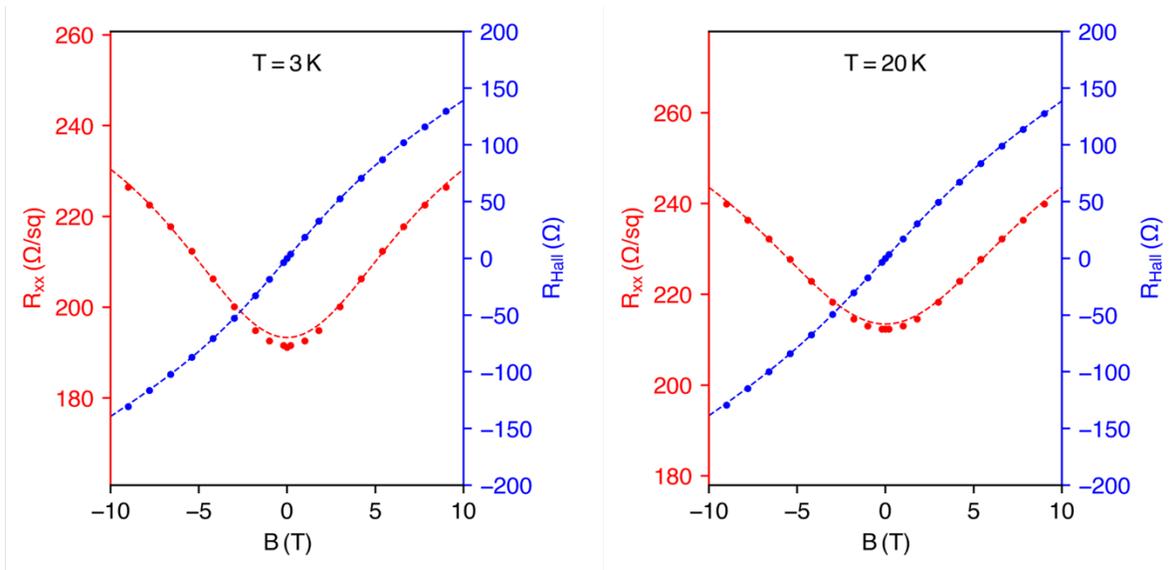

**Extended Data Fig. 2| Low-field magnetotransport measurement.** Longitudinal resistance ($R_{xx}$) and Hall resistance ($R_{\text{Hall}}$) up to 9 T were measured in a PPMS system using DC current and static magnetic fields for $T = 3$ K (left) and $T = 20$ K (right). Solid circles are data, and the dashed lines are fits to a two-channel classical Drude model[36].

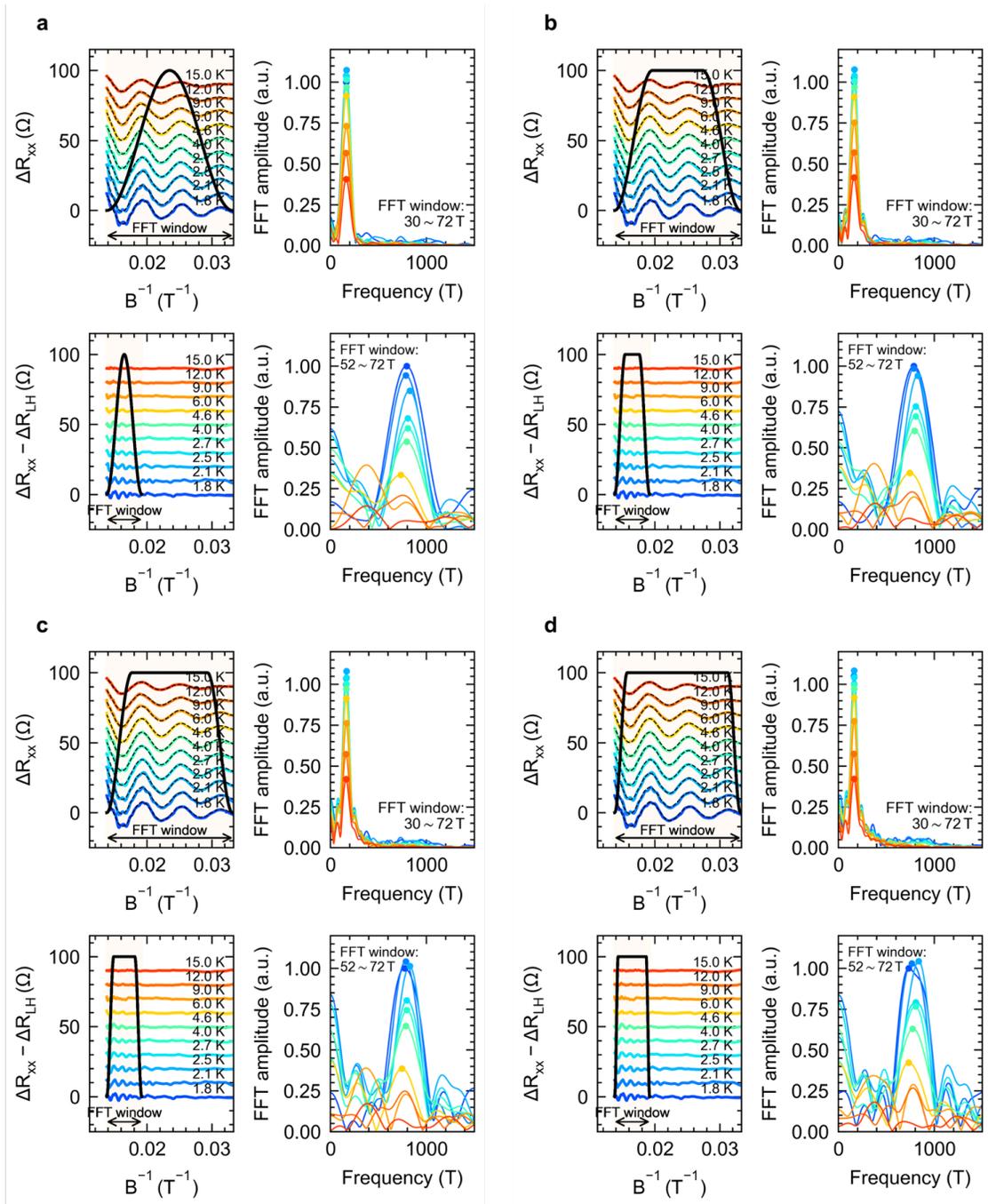

**Extended Data Fig. 3| FFT analysis using different window functions. a-d**, The background-subtracted curves oscillations (top left) are multiplied by a window function (black solid line) which tapers to zero at the minimum and maximum of the magnetic field range. The same is done for the heavy hole oscillations. The windows are Tukey windows of different values of the "$r$"-parameter: (a) $r = 1.0$ (equivalent to the Hanning window); (b) $r = 0.6$; (c) $r = 0.4$; (d) $r = 0.2$.

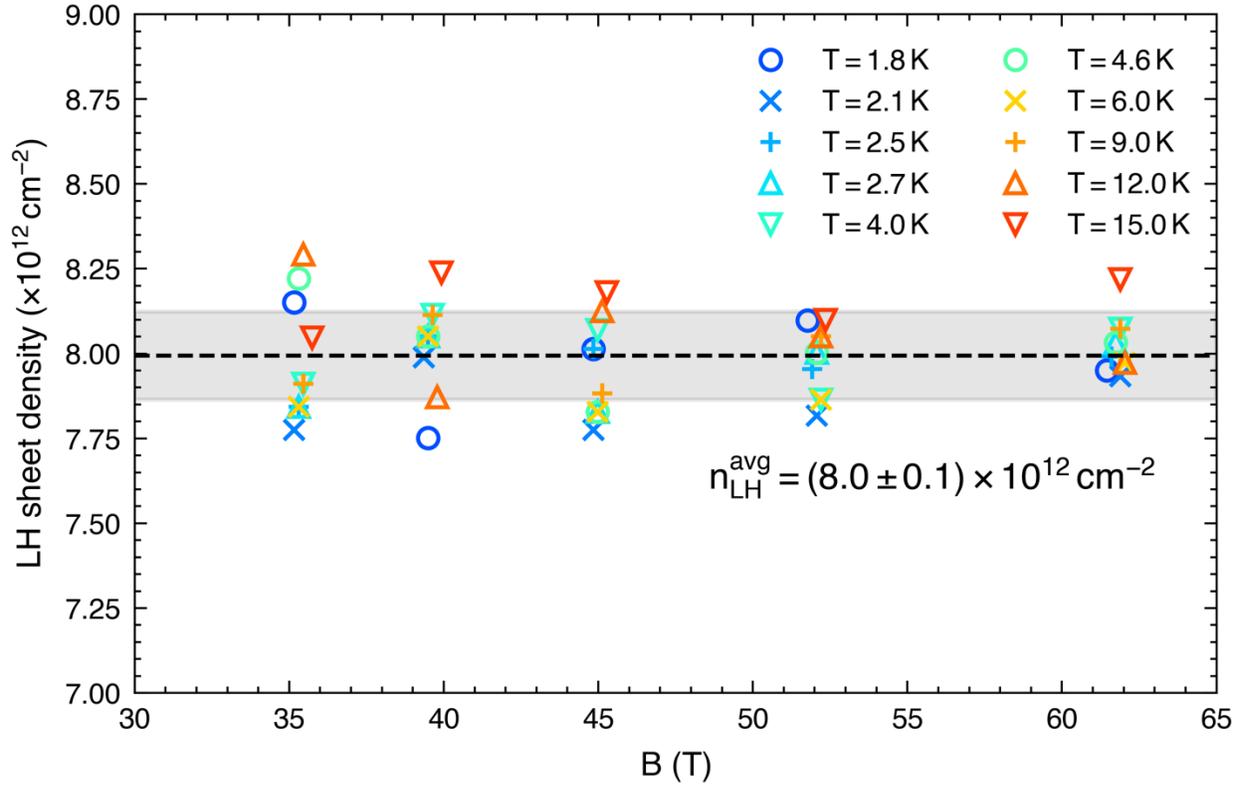

**Extended Data Fig. 4| Field independence of light hole sheet density.** The oscillation frequency at each extremum is obtained from the spacing, in $1/B$, to the previous extremum, which gives half the oscillation period $\frac{1}{2}T_{B^{-1}}$. The sheet density is then calculated in the usual way: $n = \frac{2e}{h}f = \frac{2e}{h}\frac{1}{T_{B^{-1}}}$. This procedure is applied up to the $B = 62\,T$ peak, the last fully resolved light hole oscillation. No field dependence of the light hole density is observed, consistent with the narrow width of the FFT peak in Fig. 2b and Extended Data Fig. 3. The dashed line shows the density averaged across all fields and temperatures, and the shaded region indicates one standard deviation above and below the average.

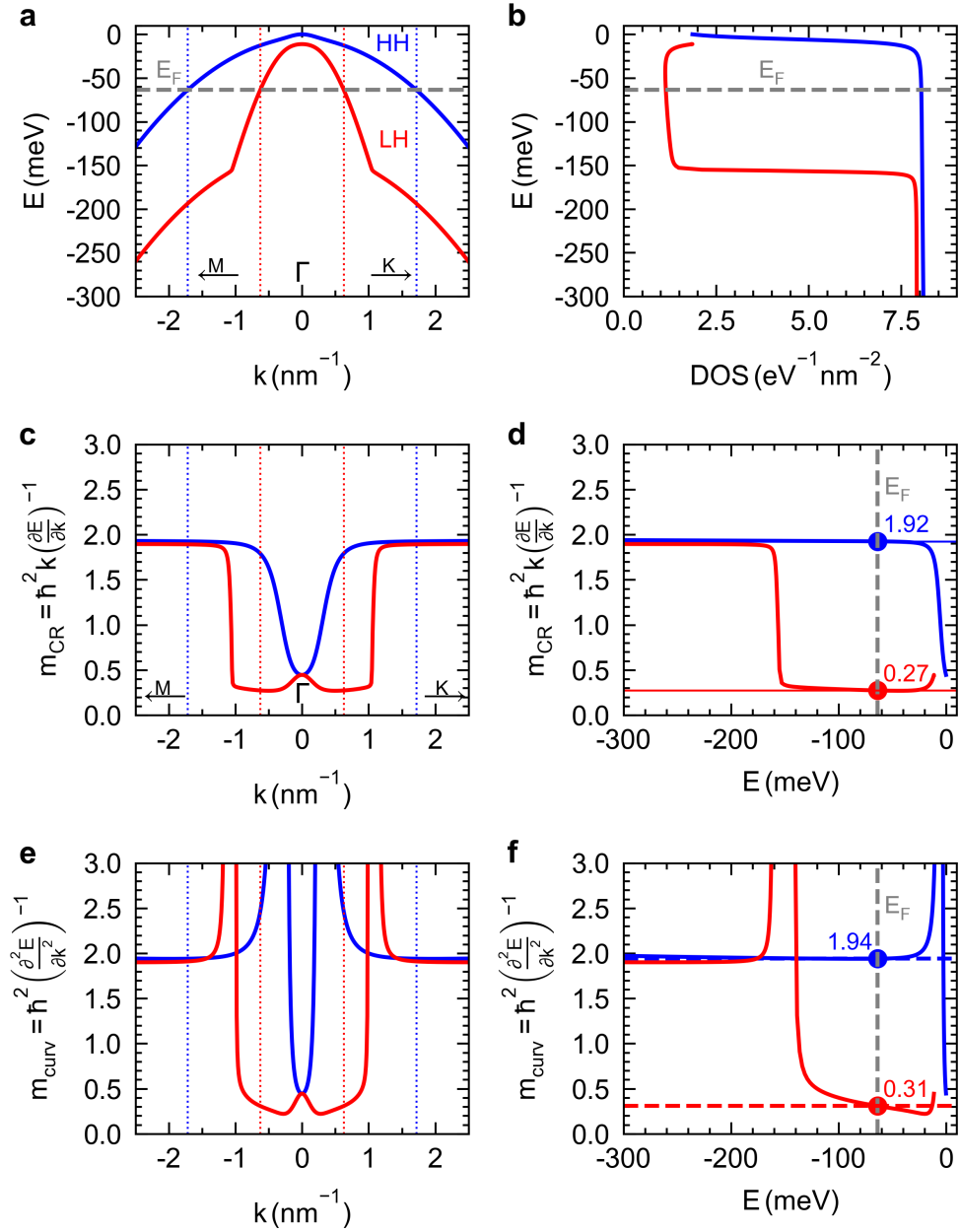

**Extended Data Fig. 5| Comparison of cyclotron effective mass and effective mass for nonparabolic dispersions. a**, In-plane dispersion of the light holes (magenta) and heavy holes (purple) around the Γ point along the M ($k < 0$) and K ($k > 0$) directions, calculated using the k.p method. The horizontal dashed line indicates the Fermi level, whose intersections with the bands give the Fermi wave vectors (vertical dashed lines) for the two bands. **b**, Density of states as a function of energy for the light and heavy holes. **c-d,** The calculated cyclotron effective mass as a function of wave vector (**c**) and energy (**d**). The values at the Fermi level (vertical dashed line in d) are shown. **e-f,** The calculated "curvature" effective mass as a function of wave vector (**e**) and energy (**f**).

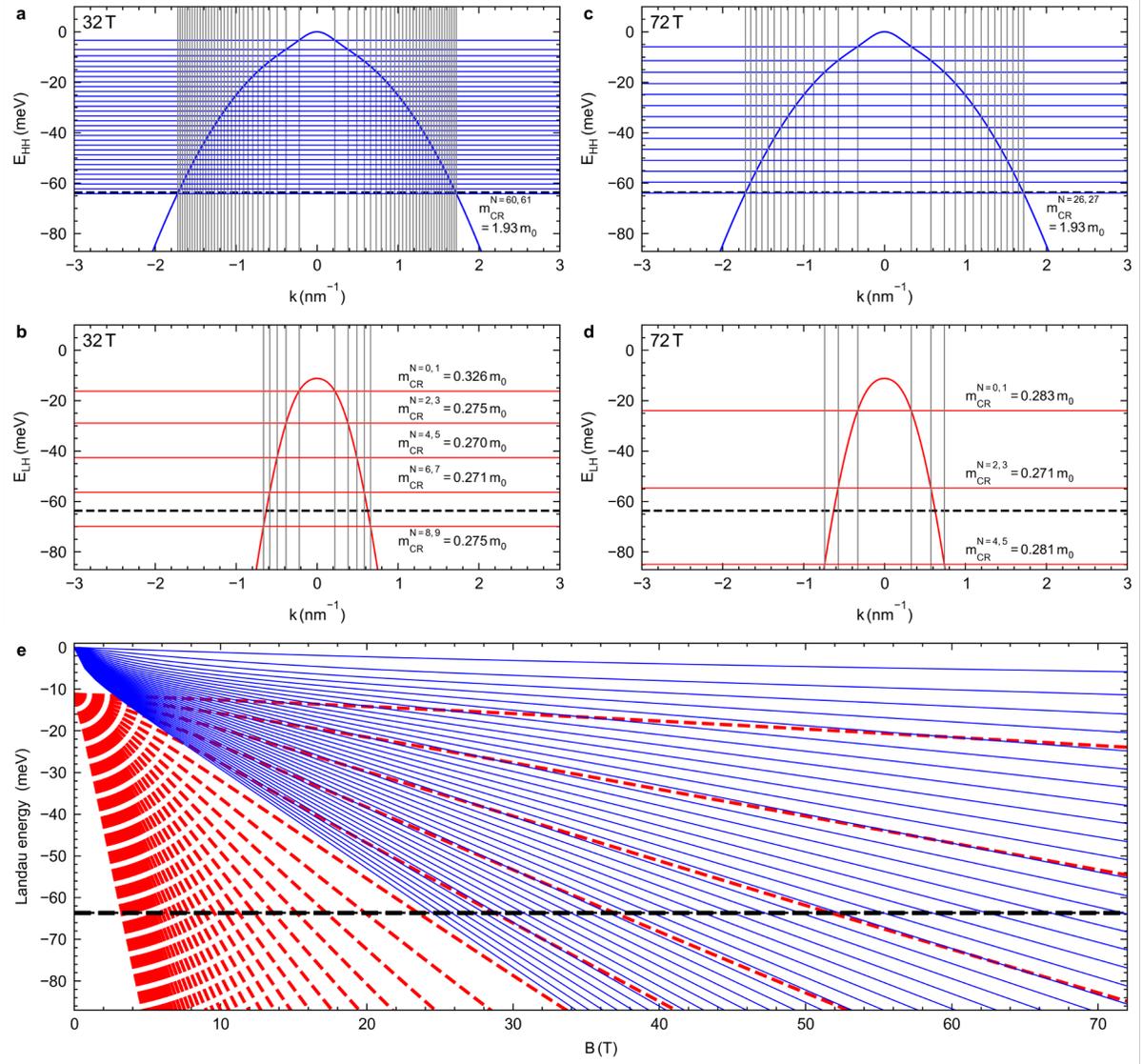

**Extended Data Fig. 6| Landau level calculations. a-d**, The quantized wave vectors are calculated for $B = 32$ T (**a, b**) and $B = 72$ T (**c, d**) and shown as vertical lines. Their intersections with the heavy hole (**a, c**) and light hole (**b, d**) dispersions determine the LLs. Only the occupied LLs and the lowest unoccupied LL are shown in (**a-d**). **e**, The Landau fan diagram showing the LLs for heavy hole (solid purple lines) and light hole (dashed magenta lines). Despite band nonparabolicity, both light hole and heavy hole LLs are approximately uniformly spaced near $E_F$. Consistently, the cyclotron masses of the LLs near $E_F$ converge to $m_{CR}^{HH} = 1.93\ m_0$ and $m_{CR}^{LH} = 0.27\sim0.28\ m_0$ at both fields.

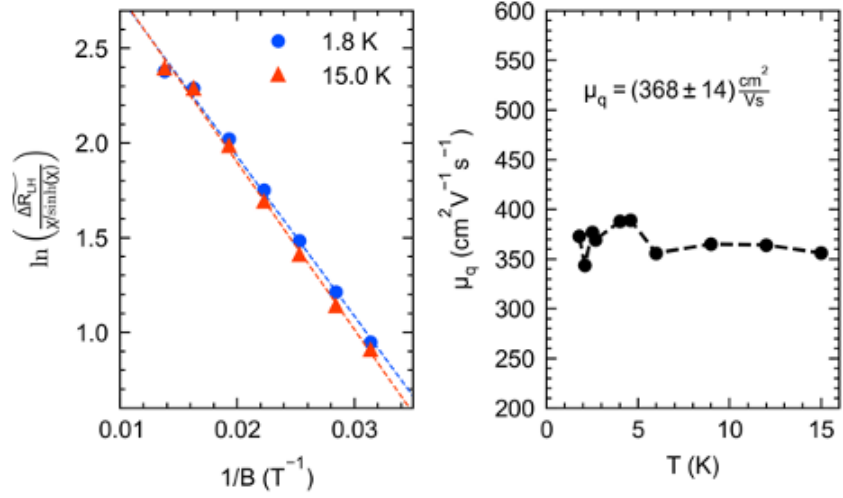

**Extended Data Fig. 7| Extraction of quantum mobility from Dingle Plot.** (Left) The Dingle plot ($\ln\left[\frac{\widetilde{\Delta R_{LH}}}{X/\sinh X}\right]$ vs. $1/B$ for the light hole peaks at $T = 1.8$ K and $T = 15$ K. $\widetilde{\Delta R_{LH}} = \frac{\Delta R_{LH}}{R^{bg}(T)}$ is the background-normalized light hole oscillation amplitude. for A linear fit (dashed line, with slope $s$) gives the quantum mobility as $\mu_q = -\pi/s$. (Right) $\mu_q$ as a function of temperature. $\mu_q$ has no discernable temperature dependence, with a temperature-averaged value of 368 cm$^2$V$^{-1}$s$^{-1}$ and a standard deviation of 14 cm$^2$V$^{-1}$s$^{-1}$.

| Parameter | GaN | AlN |
|---|---|---|
| $a^{300\,K}$ (Å) | 3.189 | 3.112 |
| $c^{300\,K}$ (Å) | 5.185 | 4.982 |
| $E_g$ (eV) | 3.510 | 6.25 |
| $\alpha$ (meV/K) | 0.909 | 1.799 |
| $\beta$ (K) | 830 | 1462 |
| $\Delta_{CR}$ (eV) | 0.010 | -0.169 |
| $\Delta_{SO}$ (eV) | 0.017 | 0.019 |
| $A_1$ | -5.947 | -3.991 |
| $A_2$ | -0.528 | -0.311 |
| $A_3$ | 5.414 | 3.671 |
| $A_4$ | -2.512 | -1.147 |
| $A_5$ | -2.510 | -1.329 |
| $A_6$ | -3.202 | -1.952 |
| $a_{cz}$ (eV) | -11.3 | -11.8 |
| $a_{ct}$ (eV) | -4.9 | -3.4 |
| $a_{cz} - D_1$ (eV) | -6.07 | -4.36 |
| $a_{ct} - D_2$ (eV) | -8.88 | -12.35 |
| $D_3$ (eV) | 5.38 | 9.17 |
| $D_4$ (eV) | -2.69 | -3.72 |
| $D_5$ (eV) | -2.56 | -2.93 |
| $D_6$ (eV) | -3.88 | -4.58 |
| $C_{11}$ (GPa) | 390 | 396 |
| $C_{12}$ (GPa) | 145 | 137 |
| $C_{13}$ (GPa) | 106 | 108 |
| $C_{33}$ (GPa) | 398 | 373 |
| $C_{44}$ (GPa) | 105 | 116 |
| $P_{sp}$ (C/m$^2$) | 1.327 | 1.341 |
| $e_{33}$ (C/m$^2$) | 1.02 | 1.569 |
| $e_{31}$ (C/m$^2$) | -1.863 | -2.027 |
| $e_{51}$ (C/m$^2$) | 0.3255 | 0.4176 |

**Extended Data Table 1|** Material parameters used for k.p and band diagram calculations shown in Fig. 4, Extended Data Fig. 5, and Extended Data Fig. 6.